\documentclass[11pt]{article}
\usepackage{graphicx}
\usepackage{amsmath}
\usepackage{amssymb}
\usepackage{amsxtra}

\title{Predictions of Additional Baryons and Mesons}
\author{Paul H. Frampton\\ \\Dipartimento di Matematica e Fisica "Ennio De Giorgi",\\
Universit\`{a} del Salento, Via Arnesano, 73100 Lecce, Italy.\\ \\paul.h.frampton@gmail.com\\}

\begin{document}
\maketitle
\begin{center}
To appear in the Proceedings of the 25th Bled Workshop "What Comes Beyond
the Standard Model". July 3-10, 2022. 
\end{center}

\begin{abstract}
\noindent
We discuss the predictions of the bilepton model which is an extension
of the standard model in which the group
$SU(2) \times U(1)$ is changed to $SU(3)\times U(1)$ and the fermion
families are treated non-sequentially with the third 
assigned differently from the first two.
Cancellation of triangle anomalies and asymptotic
freedom require three families. The predicted new physics
includes bileptons and three heavy quarks  
${\cal D}$. ${\cal S}$ and ${\cal T}$.  QCD will bind the heavy quarks
to light quarks and to each other to form baryons and mesons
which, unlike bileptons, are beyond the reach of the 
LHC but accessible in a
hypothetical 100 TeV proton-proton collider.
\end{abstract}


\section{Introduction}

\noindent
In this talk we shall discuss what now seems likely to be the first
new particle beyond the standard model and which is 
now being actively searched for at the LHC.
The bilepton model, a better name than the 331-Model, was invented as an example 
of what then was expected to be a new class of models which require the existence
three families. That invention was in 1992 \cite{331} but
it required a couple more years to realise that the expected new class of
models has only one member.

\bigskip

\noindent
We remain optimistic that LHC can find a discovery signal for the
bilepton gauge boson in the remainder of 2022. 
What we can say generally is that to invent a model which is beyond the standard 
model, one generally aims to both  (i)address and solve a question
unanswered within the standard model, and to (ii) provide explicit predictions
which are testable.  The bilepton model beautifully fulfils both of
these criteria.

\bigskip

\noindent
Although not a sub-theory, the model originated from studying an interesting
$SU(15)$ model\cite{SU(15)} in which the 224 gauge bosons couple
to all possible pairs of the 15 states
\begin{equation}
(u, d)_{\alpha}; (\bar{u})_{\alpha}; (\bar{d})_{\alpha}; (\nu_e. e); (\bar{e}).
\label{fifteen}
\end{equation}
Every gauge boson therefore has a well-defined B and L so there 
can be no proton decay by tree-level gauge boson exchange.

\bigskip

\noindent
In the $SU(15)$ model there is one unaesthetic feature that
anomalies are cancelled by adding mirror fermions as in $15 + \bar{15}$.
But persisting further, we considered the subgroups in
$SU(15) \rightarrow SU(12)_q \times SU(3)_l$, especially the
$SU(3)_l$ which contains an antitriplet $(e^+,\nu_e, e^-)$
where the $|L|=|Q|=2$ bilepton can first be seen, coupling electron to positron.

\bigskip

\noindent
The question then was: can a chiral model contain bileptons?
After hundreds of  trials and errors we found only
one solution of the anomaly cancellation equations.
This required non-sequential families where the third is
assigned differently from the first two and explains why there must be
three families. This is the bilepton model. It provides an answer to
Rabi's famous question when the muon was discovered in 1936:
"Who ordered that?"

\bigskip

\noindent
The non-sequentiality of families offers one explanation for
the failure of the $SU(5)$ model studied first in 1974
\cite{GG,GQW} then in hundreds of other papers.
$SU(5)$
assumed sequentiality of
families of the form $3(10 + \bar{5})$. 

\bigskip

\noindent
In 1977 Weinberg\cite{LW}
and in 1984 Glashow \cite{G} both considered upgrading 
the electroweak $SU(2)$ of the standard model
to $SU(3)$ but overlooked the assignments 
which explain three families.

\section{Bilepton model}

\noindent
The gauge group is:
\begin{equation}
SU(3)_C \times SU(2)_L \times U(1)_X
\end{equation}
The simplest choice for the electric charge is
\begin{equation}
Q = \frac{1}{2} \lambda_L^3 + \left( \frac{\sqrt{3}}{2} \right) \lambda_L^8  + X \left(
 \frac{\sqrt{3}}{\sqrt{2}} \right) \lambda^9
 \end{equation}
 where
 \begin{equation}
 Tr (\lambda_L^a \lambda_L^b) = 2  \delta^{ab}
 \end{equation}
 and
 \begin{equation}
 \lambda^9 \equiv \left( \frac{\sqrt{2}}{\sqrt{3}} \right) diag (1, 1, 1)
 \end{equation}
 
 \bigskip
 
\noindent
Thus a triplet has charges $(X+1, X, X-1)$.

\bigskip

\noindent
Leptons are treated democratically in each \\
of the three families. They are colour singlets in antitriplets of $SU(3)_L$ :

\bigskip

$(  e^+, \nu_e, e^- )_L$

\bigskip

$( \mu^+, \nu_{\mu}, \mu^- )_L$

\bigskip

$( \tau^+, \nu_{\tau}, \tau^- )_L$

\bigskip

\noindent
All have $X=0$.

\bigskip

\noindent
Quarks in the first family are assigned to a left-handed triplet plus three singlets of $SU(3)_L$.

\bigskip

$( u^{\alpha}, d^{\alpha}, D^{\alpha} )_L~~~ (\bar{u}_{\alpha})_L,  (\bar{d}_{\alpha})_L, 
(\bar{D}_{\alpha})_L$

\bigskip

\noindent
Similarly for the second family

\bigskip

$( c^{\alpha}, s^{\alpha}, S^{\alpha} )_L~~~ (\bar{c}_{\alpha})_L,  (\bar{s}_{\alpha})_L, 
(\bar{S}_{\alpha})_L$

\bigskip

\noindent
The X values are for the triplets are $X=-1/3$ and for the singlets
$X=-2/3, +1/3, +4/3$ respectively. The electric charge of the new
quarks $D, S$  is $-4/3$.

\bigskip

\noindent
The quarks of the third family are treated differently. They are assigned
to a left-handed antitriplet and three singlets under $SU(3)_L$

\bigskip

$( b^{\alpha}, t^{\alpha}, T^{\alpha} )_L~~~ (\bar{b}_{\alpha})_L,  (\bar{t}_{\alpha})_L, 
(\bar{T}_{\alpha})_L$

\bigskip

\noindent
The antitriplet has $X=+2/3$ and the singlets carry $X=+1/3, -2/3, -5/3$ respectively. 
The new quark $T$ has $Q=5/3$.

\bigskip

\noindent
Some of the relevant LHC phenomenology is discussed
in \cite{CCCF}. A refined mass estimate \cite{CF} for the bilepton is
is $M(Y^{\pm\pm}) = (1.29 \pm 0.06)$ TeV where {\it faute de mieux} it was assumed that the
symmetry breaking of $SU(3)_L$ is closely similar to that of $SU(2)_L$. It will be pleasing
if the physical mass is consistent with this.

\bigskip

\noindent
\section{New Quarks}

\bigskip

\noindent
Because the quarks are in triplets and anti-triplets of $SU(3)_L$, rather than only in doublets of $SU(2)_L$ as in the
standard model, there is necessarily an additional quark in each family. In the first and second families they are the
${\cal D}$ and ${\cal S}$ respectively, both with charge $Q=-4/3$ and lepton number $L=+2$. In the third family 
is the ${\cal T}$
with charge $Q=+5/3$ and lepton number $L=-2$. All the three TeV scale quarks are colour triplets with
spin-$\frac{1}{2}$ and baryon number $B=\frac{1}{3}$. Their masses
are yet to be measured but may be expected to be below the ceiling of $4.1 TeV$ which is the upper limit
for symmetry breaking of $SU(3)_L$ and probably above $1 TeV$. By analogy with the
known quarks, one might expect $M({\cal T}) > M({\cal S}) > M({\cal D})$, although 
without experimental data this is
conjecture.

\bigskip

\noindent
The heavy quarks and antiquarks will be bound to light quarks and antiquarks, and to each other, to form an interesting spectroscopy of
mesons and baryons. Let us first display, in Tables 1, 2 the TeV mesons, then in Tables 3,4,5 the TeV baryons. The charge conjugate
states are equally expected, and will reverse the signs of $Q$ and $L$.

\bigskip

\noindent
\section{Additional Baryons and Mesons}.

\bigskip

\begin{table}[h]
\caption{TeV mesons ${\cal Q}\bar{q}$}
\begin{center}
\begin{tabular}{||c|c||c|c||}
\hline
\hline
${\cal Q}$ & $\bar{q}$ & Q & L \\
\hline
\hline
&&& \\
${\cal D/S}$ & $\bar{u}$ etc. & -2 & +2 \\
 ${\cal D/S}$ & $\bar{d}$ etc. & -1 & +2 \\
 ${\cal T}$ & $\bar{u}$ etc. & +1 & -2 \\
  ${\cal T}$ & $\bar{d}$ etc.& +2 & -2 \\
  &&& \\
\hline
\hline
\end{tabular}
\end{center}
\label{mesonsQq}
\end{table}

\bigskip

\begin{table}[t]
\caption{TeV mesons ${\cal Q} \bar{{\cal Q}}$}
\begin{center}
\begin{tabular}{||c|c||c|c||}
\hline
\hline
& & & \\
 ${\cal Q}$ & $\bar{{\cal Q}}$ & Q & L \\
\hline
&&&\\
${\cal D/S}$ & ${\cal \bar{D}/\bar{S}}$ & 0 & 0 \\
${\cal D/S}$ & $\bar{{\cal T}}$ &  -3 & +4 \\
${\cal T}$ & $\bar{T}$ & 0  & 0 \\
&&& \\
\hline
\hline
\end{tabular}
\end{center}
\label{mesonsQq}
\end{table}

\bigskip

\begin{table}[h]
\caption{TeV baryons ${\cal Q}qq$}
\begin{center}
\begin{tabular}{||c|c||c|c||}
\hline
\hline
&&& \\
 ${\cal Q}$ & qq & Q & L \\
\hline
&&&\\
${\cal D/S}$ & dd etc. & -2 & +2 \\
${\cal D/S}$ & ud etc. & -1 & +2 \\
${\cal D/S}$ & uu etc. & 0  & +2 \\
${\cal T}$ & dd etc. & +1 & -2 \\
${\cal T}$ & ud etc. & +2 & -2 \\
${\cal T}$ & uu etc, & +3 & -2 \\ 
&&& \\
\hline
\hline
\end{tabular}
\end{center}
\label{mesonsQq}
\end{table}

\bigskip

\begin{table}[t]
\caption{TeV baryons ${\cal Q}{\cal Q}q$}
\begin{center}
\begin{tabular}{||c|c||c|c||}
\hline
\hline
& & & \\
 ${\cal Q}{\cal Q} $ & $q$ & Q & L \\
\hline
&&&\\
${\cal (D/S)(D/S)}$ & d etc. & -3 & +4 \\
${\cal (D/S)(D/S)}$ & u etc.& -2  & +4 \\
${\cal (D/S)T}$ & d etc. & 0 & 0 \\
${\cal (D/S)T}$ & u etc.&+1 & 0  \\
${\cal TT}$ & d etc. & +3 & -4 \\
${\cal TT}$ & u etc. & +4 & -4 \\ 
&&& \\
\hline
\hline
\end{tabular}
\end{center}
\label{mesonsQq}
\end{table}

\begin{table}[h]
\caption{TeV baryons ${\cal Q}{\cal Q}{\cal Q}$}
\begin{center}
\begin{tabular}{||c||c|c||}
\hline
\hline
& &  \\
 ${\cal Q}{\cal Q}{\cal Q}$ & Q & L \\
\hline
&&\\
${\cal (D/S)(D/S)(D/S)}$ & -4 & +6 \\
${\cal (D/S)(D/S)T}$ & -1  & +2 \\
${\cal (D/S)TT}$ & +2 & -2 \\
${\cal TTT}$ & +5 & -6  \\ 
&& \\
\hline
\hline
\end{tabular}
\end{center}
\label{mesonsQq}
\end{table}

\bigskip

\noindent
Although the ${\cal Q}$ masses are unknown, it may be reasonable first
to make a preliminary discussion of these states by assuming that
\begin{equation}
M({\cal T})> M({\cal S}) + 2M_t > M({\cal D}) +4M_t
\label{simplification}
\end{equation}

\noindent
where $M_t$ is the top quark mass so that the lightest of the TeV baryons and mesons
are those containing just one ${\cal D}$ quark or one ${\cal \bar{D}}$ antiquark. The next lightest are
the TeV baryons and mesons containing just one ${\cal S}$ quark or one $\bar{{\cal S}}$ antiquark.

\bigskip

\noindent
We begin by discussing the decay modes of the ${\cal D}\bar{q}$
mesons in Table 1, focusing on final states from the first family.
The decays of ${\cal D}$ include, taking care of $L$ conservation,

\begin{eqnarray}
{\cal D} &\rightarrow& d + Y^- \nonumber \\
& \rightarrow & d + (e^- + \nu_e) \nonumber \\
& \rightarrow & d + (\mu^- + \nu_{\mu}) \nonumber \\
& \rightarrow & d + (\tau^- + \nu_{\tau}) \nonumber \\
\label{calDdecay}
\end{eqnarray}
which implies that decays of the $({\cal D} \bar{u})$ meson include
\begin{eqnarray}
({\cal D} \bar{u}) & \rightarrow & \pi^- + (e^- + \nu_e) \nonumber \\
& \rightarrow & \pi^- + (\mu^- + \nu_{\mu}) \nonumber \\
& \rightarrow &  \pi^- + (\tau^- + \nu_{\tau}) \nonumber \\
\label{calDudecay}
\end{eqnarray}
and variants thereof where $\pi^-$ is replaced by any other non-strange
negatively charged meson. The $d$ in Eq.(\ref{calDdecay}) can be
replaced by $s$ or $b$ which subsequently decay.

\bigskip

\noindent
An alternative to Eq.(\ref{calDdecay}) is
\begin{eqnarray}
{\cal D} &\rightarrow& u + Y^{- -} \nonumber \\
& \rightarrow & u + (e^- + e^-) \nonumber \\
& \rightarrow & u + (\mu^- + \mu^-) \nonumber \\
& \rightarrow & u + (\tau^- + \tau^-) \nonumber \\
\label{calDdecay2}
\end{eqnarray}
which implies additional decay modes of the $({\cal D} \bar{u})$ meson which include
\begin{eqnarray}
({\cal D} \bar{u}) & \rightarrow & \pi^0 + (e^- + e^-) \nonumber \\
& \rightarrow & \pi^0 + (\mu^- +\mu^-) \nonumber \\
& \rightarrow &  \pi^0 + (\tau^- + \tau^-) \nonumber \\
\label{calDudecay2}
\end{eqnarray}
and variants obtained by flavour replacements.  Eqs.(\ref{calDudecay}) and (\ref{calDudecay2}),
and their generalisations to other flavours,
suffice to illustrate the richness of $({\cal D}\bar{u})$ decays.

\bigskip

\noindent
Turning to the meson ${\cal D} \bar{d}$, we can use
Eq.(\ref{calDdecay}) to identify amongst its possible decays

\begin{eqnarray}
({\cal D} \bar{d}) & \rightarrow & \pi^0 + (e^- + \nu_e) \nonumber \\
& \rightarrow & \pi^0 + (\mu^- + \nu_{\mu}) \nonumber \\
& \rightarrow &  \pi^0 + (\tau^- + \nu_{\tau}) \nonumber \\
\label{calDddecay}
\end{eqnarray}
and variants thereof where $\pi^0$ is replaced by any other non-strange
neutral meson. When $u$ in Eq.(\ref{calDdecay}) is
replaced by $c$ or $t$ which subsequently decay, we arrive
at many other decay channels additional to Eq.(\ref{calDddecay}).

\bigskip

\noindent
Employing instead the ${\cal D}$ decays in Eq.(\ref{calDdecay2})
 implies additional decay modes of $({\cal D} \bar{d})$ meson that include
\begin{eqnarray}
({\cal D} \bar{d}) & \rightarrow & \pi^+ + (e^- + e^-) \nonumber \\
& \rightarrow & \pi^+ + (\mu^- +\mu^-) \nonumber \\
& \rightarrow &  \pi^+ + (\tau^- + \tau^-) \nonumber \\
\label{calDddecay2}
\end{eqnarray}
and variants obtained by flavour replacement.  Eqs.(\ref{calDddecay}) and (\ref{calDddecay2}),
merely illustrate a few of the simplest $({\cal D}\bar{d})$ decays. There are many more.

\bigskip

\noindent
Next we consider the lightest TeV baryons in Table 3 with ${\cal Q} = {\cal D}$.
Using the ${\cal D}$ decays from Eq.(\ref{calDdecay}) we find
for $({\cal D}uu)$ decay
\begin{eqnarray}
({\cal D}uu) & \rightarrow & p + (l_i^- + \nu_i). \nonumber \\
  \label{DuuDecay}
  \end{eqnarray}
together with flavour rearrangements. Here, as in subsequent equations,
$i=e,\mu,\tau$.

\bigskip

\noindent
Alternatively, the ${\cal D}$ decays from Eq.(\ref{calDdecay2})
lead to
\begin{eqnarray}
({\cal D}uu) & \rightarrow & N^{*++} + Y^{- -} \nonumber.  \\
&\rightarrow & p + \pi^+ + (l_i^- + l_i^-). \nonumber.  \\
\label{DuuDecay2}
\end{eqnarray}

\noindent
Looking at the TeV baryon $({\cal D} ud)$ the respective
sets of decays corresponding to Eq.(\ref{calDdecay}) are
\begin{eqnarray}
({\cal D}ud) & \rightarrow & n +(l_i^- + \nu_i) \nonumber \\
\label{DudDecay}
\end{eqnarray}
where only the simplest light baryon is exhibited.

\bigskip

\noindent
Corresponding to ${\cal D}$ decays in Eq.(\ref{calDdecay2})
there are also
\begin{eqnarray}
({\cal D}ud) & \rightarrow & p +(l_i^- + l_i^-) \nonumber \\
\label{DudDecay}
\end{eqnarray}
in the simplest cases.

\bigskip

\noindent
Finally, of the $({\cal D}qq)$ TeV baryons, we write out the
decays for $({\cal D}dd)$, first for the ${\cal D}$ decays
in Eq.(\ref{calDdecay})
\begin{eqnarray}
({\cal D}dd) & \rightarrow & N^{*-} + Y^- \nonumber \\
& \rightarrow & n + \pi^- + (l_i^- + \nu_i). \nonumber \\
\label{DddDecay}
\end{eqnarray}
within flavour variations.

\bigskip

\noindent
With the Eq.(\ref{calDdecay2}) decays of ${\cal D}$ there are also decays
\begin{eqnarray}
({\cal D}dd) & \rightarrow & n +(l_i^- + l_i^-) \nonumber \\
\label{DudDecay}
\end{eqnarray}
again with more possibilities by choosing alternative flavours.

\bigskip

\noindent
We now replace the TeV quark ${\cal D}$ by the next heavier
TeV quark ${\cal S}$ and repeat our study of decays whereupon we shall
encounter the first example of decay not only to the known
quarks but also to a TeV quark.

\bigskip

\noindent
The TeV quark ${\cal S}$ has possible decay channels

\begin{eqnarray}
{\cal S} &\rightarrow& d + Y^- \nonumber \\
& \rightarrow & d + (e^- + \nu_e) \nonumber \\
& \rightarrow & d + (\mu^- + \nu_{\mu})   \nonumber \\
& \rightarrow & d + (\tau^- + \nu_{\tau})     \nonumber \\
& \rightarrow &         {\cal D}   + Z'               \nonumber \\
& \rightarrow & d + (e^- + \nu_e) + (e^+ + e^-)  \nonumber \\
& \rightarrow & d + (e^- + \nu_e) + (\mu^+ + \mu^-)  \nonumber \\
& \rightarrow & d + (e^- + \nu_e) + (\tau^+ + \tau^-)  \nonumber \\
& \rightarrow & d + (\mu^- + \nu_{\mu}) + (e^+ + e^-)  \nonumber \\
& \rightarrow & d + (\mu^- + \nu_{\mu}) + (\mu^+ + \mu^-)  \nonumber \\
& \rightarrow & d + (\mu^- + \nu_{\mu}) + (\tau^+ + \tau^-)  \nonumber \\
& \rightarrow & d + (\tau^- + \nu_{\tau}) + (e^+ + e^-) \nonumber \\
& \rightarrow & d + (\tau^- + \nu_{\tau}) + (\mu^+ + \mu^-) \nonumber \\
& \rightarrow & d + (\tau^- + \nu_{\tau}) + (\tau^+ + \tau^-) \nonumber \\
\label{calSdecay}
\end{eqnarray}
where we note the opening up of channels due to 
${\cal S}\rightarrow{\cal D}$ decay.

\bigskip

\noindent
With Eq.(\ref{calSdecay}) in mind, the decays
of the TeV meson $({\cal S} \bar{u})$ include
\begin{eqnarray}
({\cal S} \bar{u}) & \rightarrow & \pi^- + (l_i^- + \nu_i) \nonumber \\
& \rightarrow & \pi^- + (l_i^- + \nu_i) + (l_j^+ + l_j^-)  \nonumber \\
\label{calSudecay}
\end{eqnarray}
where the second line involves a ${\cal D}$
intermediary.

\bigskip

\noindent
An alternative to Eq.(\ref{calSdecay}) is
\begin{eqnarray}
{\cal S} &\rightarrow& u + Y^{- -} \nonumber \\
& \rightarrow & u + (e^- + e^-) \nonumber \\
& \rightarrow & u + (\mu^- + \mu^-) \nonumber \\
& \rightarrow & u + (\tau^- + \tau^-) \nonumber \\
\label{calSdecay2}
\end{eqnarray}
which implies additional decay modes of $({\cal S} \bar{u})$ 
\begin{eqnarray}
({\cal S} \bar{u}) & \rightarrow & \pi^0 + (l_i^- + l_i^-) \nonumber \\
\label{calSudecay2}
\end{eqnarray}
and variants which replace $\pi^0$ by another neutral non-strange meson.
Eqs.(\ref{calSudecay}) and (\ref{calSudecay2}),
illustrate sufficiently $({\cal S}\bar{u})$ decays.

\bigskip

\noindent
Turning to the meson (${\cal S} \bar{d}$), we can use
Eq.(\ref{calSdecay}) to identify its possible decays
\begin{eqnarray}
({\cal S} \bar{d}) & \rightarrow & \pi^0 + (l_i^- + \nu_i) \nonumber \\
\label{calSddecay}
\end{eqnarray}
When $u$ in Eq.(\ref{calSdecay}) is
replaced by $c$ or $t$ which subsequently decay, we arrive
at many other decay channels additional to Eq.(\ref{calSddecay}).

\bigskip

\noindent
Employing instead the ${\cal S}$ decays in Eq.(\ref{calSdecay2})
implies additional decay modes of $({\cal S} \bar{d})$ that include

\begin{eqnarray}
({\cal S} \bar{d}) & \rightarrow & \pi^+ + (l_i^- + l_i^-) \nonumber \\
\label{calSddecay2}
\end{eqnarray}
and variants obtained by flavour replacement.  Eqs.(\ref{calSddecay}) and (\ref{calSddecay2}),
illustrate only a few of the simplest $({\cal S}\bar{d})$ decays. There are many more.

\bigskip

\noindent
Next we consider the lightest TeV baryons in Table 3 with one ${\cal Q} = {\cal S}$.
Using the ${\cal S}$ decays from Eq.(\ref{calSdecay}) we find
for (${\cal S}uu$) decay
\begin{eqnarray}
({\cal S}uu) & \rightarrow & p + (l_i^- + \nu_i). \nonumber \\
  \label{SuuDecay}
  \end{eqnarray}
\noindent
together with flavour rearrangements.

\bigskip

\noindent
Alternatively, the ${\cal S}$ decays from Eq.(\ref{calSdecay2})
lead to
\begin{eqnarray}
({\cal S}uu) & \rightarrow & N^{*++} + (l_i^- + l_i^-). \nonumber.  \\
& \rightarrow & p + \pi^+ + (l_i^- + l_i^-). \nonumber \\
\label{SuuDecay2}
\end{eqnarray}

\bigskip

\noindent
Looking at the TeV baryon $({\cal S} ud)$ the respective
sets of decays corresponding to Eq.(\ref{calSdecay}) are
\begin{eqnarray}
({\cal S}ud) & \rightarrow & n +(l_i^- + \nu_i) \nonumber \\
\label{SudDecay}
\end{eqnarray}
where only the simplest version is exhibited.

\bigskip

\noindent
Corresponding to the ${\cal S}$ decays in Eq.(\ref{calSdecay2})
there are the decays
\begin{eqnarray}
({\cal S}ud) & \rightarrow & p +(l_i^- + l_i^-) \nonumber \\
\label{SudDecay}
\end{eqnarray}

\bigskip

\noindent
For baryon $({\cal S}dd)$, firstly from the ${\cal S}$ decays
in Eq.(\ref{calSdecay}) we have
\begin{eqnarray}
({\cal S}dd) & \rightarrow & N^{*-} + Y^- \nonumber \\
& \rightarrow & n + \pi^- + (l_i^- + \nu_i). \nonumber \\
\label{SddDecay}
\end{eqnarray}
within flavour variations.

\bigskip

\noindent
Secondly, from the Eq.(\ref{calSdecay2}) decays of ${\cal S}$ there
are baryon decays of the type
\begin{eqnarray}
({\cal S}dd) & \rightarrow & n +(l_i^- + l_i^-) \nonumber \\
\label{SddDecay}
\end{eqnarray}
with more possibilities by choosing alternative flavours.

\bigskip

\noindent
\section{Discussion}

\bigskip

\noindent
We could continue further to study decays
of all the baryons and mesons in our Tables. 
However, it seems premature
to do so, until we
know from experimental data the masses and mixings
of ${\cal D,S,T}$.
We remark only that the type
of lepton cascade which we have 
exhibited in Eq.(\ref{calSdecay})
becomes ever more prevalent as the 
lepton number of the decaying hadron increases.

\bigskip

\noindent
We may expect, by analogy with the top quark mass
being close to the weak scale
 that the mass of the ${\cal T}$
quark, although probably below $4.1$ TeV for the
symmetry-breaking reason discussed {\it ut supra},
might be not much below. For example it might exceed $3$ TeV 
whereupon the mass of
a (${\cal TTT}$)  baryon could exceed $9$ TeV. 
Since this baryon has high lepton number, it must be pair
produced and such production is far beyond the reach
of the $14$ TeV LHC.  Its study would require a $100$ TeV collider 
of the type presently under preliminary discussion.  As a foretaste of the physics
accessible to such a hypothetical collider, the simplest decay of 
the $({\cal TTT})$ baryon we can find is 

\begin{equation}
p+ 4(e^+) + 2(\bar{\nu}_e).   \nonumber \\
\end{equation}
which would be 
very exciting to confirm.

\bigskip

\noindent
At the time of writing, the particles exhibited
in our Tables are conjectural. After the bilepton is discovered 
the existence of all the additional 
baryons and mesons in our  five Tables
would become a sharp predictions.

\bigskip

\noindent
The bilepton resonance in $\mu^{\pm}\mu^{\pm}$ has been the subject of searches
by the ATLAS and CMS Collaborations at the LHC, starting in March 2021. In 
March 2022, ATLAS published an inconclusive result \cite{ATLAS} about the existence
of the resonance, putting only a lower mass limit $M_Y > 1.08$ TeV. CMS has
better momentum resolution and, what is the same thing, charge identification than ATLAS and should
be able to investigate the bilepton resonance proper. The high sensitivity of CMS is a result
of serendipity because it
was designed in 1993 not for the bilepton but to search for heavy Z-primes\cite{Virdee}.
A second serendipity was an accidental 2015 meeting in London
between us and Sir Tejinder Virdee
who helped design the CMS detector. 

\bigskip

\noindent
Our strong belief in the existence of the bilepton lies partly in the 
close relationship between
the 1961 paper\cite{Glashow} which solved the parity puzzle and
our 1992 paper\cite{331} which solved the family puzzle. We regard
these two
papers which span three decades as well-matched bookends,

\bigskip

\noindent
According to our calculations \cite{CCCF}, the Run 2 data with 139/fb
collected by 2018 are sufficient for a CMS discovery of the bilepton. If not,
future LHC runs up to
their target integrated luminosity of 4/ab can provide 28 times as many events
and bilepton discovery would be merely postponed.
We do hope, however, that a great discovery will be made by the LHC within six months from
today (July 25, 2022).

\bigskip
\bigskip
\bigskip

\noindent
{\it Note added:}\\
\noindent
We answer here one interesting question received after our
talk: Why are these heavy states not
as unstable as the top quark which lives for less than
a trillion trillionth of a second? The answer is that they
decay via bilepton exchange. This fact renders their lifetimes a
trillion times longer than the top quark lifetime.

\section*{Acknowlegement}
\noindent
We thank the organisers Norma Borstnik, Maxim Khlopov and
Holger Nielsen for their invitation to present this talk.

\end{document}